\documentclass[a4paper,11pt,onecolumn,final,notitlepage,oneside]{article}
\usepackage{cmap}
\usepackage[cp1251]{inputenc}
\usepackage[T2A]{fontenc}
\usepackage[english,russian]{babel}
\usepackage{indentfirst}
\usepackage{geometry}
\usepackage[dvips]{graphicx}
\usepackage[final]{epsfig}
\usepackage{multicol}
\usepackage{graphicx}
\usepackage{subfigure}
\usepackage{amsmath}
\usepackage{enumerate}
\usepackage{amssymb}
\usepackage{amscd}
\usepackage{hhline}
\usepackage{multirow}
\usepackage{setspace}
\usepackage{makeidx}
\usepackage{textcomp}
\usepackage{amsfonts}
\usepackage{afterpage}
\usepackage{longtable}
\usepackage{cite}
\usepackage{rawfonts}
\usepackage{array}
\graphicspath{{pictures/}}

\DeclareGraphicsExtensions{.pdf,.png,.jpg,.eps}

\numberwithin{equation}{subsubsection}

\makeatletter
\def\@seccntformat#1{\csname the#1\endcsname.\ } 
\def\@biblabel#1{#1.} 
\makeatother
\addcontentsline{toc}{subsubsection}{References}
\addto\captionsrussian{}

\begin{document}
\title{\normalsize 
\begin{flushleft}
{subject:\textit{\,\,relativistic Kinematics}}
\end{flushleft}
\begin{flushleft}
{PACS: 03.30.+p}
\end{flushleft}
\vspace{\baselineskip}
\normalsize \bf THE PROPER CHARACTERISTICS OF FRAME REFERENCE AS A 4-INVARIANTS}
\author{\bf \small V. V. Voytik\,\\ 
\small \itshape teacher, Department of Medical Physics with a course of Informatics , \\ \small \itshape Bashkirian State Medical University, \\ \small \itshape Lenin st., 3, Ufa, 450008, Russia\\
\small \itshape e-mail: voytik1@yandex.ru\\
\small \itshape Received 16.02.2015 г.\\
\small \itshape Published in STFI, 2015, no.1.\\}
\date{}
\maketitle
\renewcommand{\abstractname}{}

\begin{abstract}
The paper proposes 4-dimensional equations for the proper characteristics of a rigid reference frame
\[ {{{W}'}^{\gamma }}=\Lambda ^{0}_{\;\;i}\frac{d\Lambda ^{\gamma i}}{d{t}}\,\,,\]
\[	{{{\Omega }'}^{\gamma }}=-\frac{1}{2}\,{{e}^{\alpha \mu \gamma }}\Lambda _{\;\;i}^{\mu }\frac{d{{\Lambda }^{\alpha i}}}{d{t}}\,\,.\]
 From these conditions follow the law of motion of the proper tetrad and the equations of the inverse problem of kinematics, i.e., differential equations that solve the problem of restoring the motion parameters of a rigid reference frame from known proper acceleration and angular velocity. In particular, it is shown that when boosted, a moving reference frame that has proper Thomas precession relative to the new laboratory frame will have a combination of two rotations: the new Thomas proper precession and the Wigner rotation, which together give the original frequency of Thomas precession
$$ \frac{1-\sqrt{1-v^{2}}}{v^{2}\sqrt{1-v^{2}}}\,\,e^{\beta\mu\nu}v^{\mu}\dot{v}^{\nu}=\omega'^{\beta}_{W} + b^{\beta\alpha}_{W}\,\frac{1-\sqrt{1-v^{*2}}}{v^{*2}\sqrt{1-v^{*2}}}\,\,e^{\alpha\mu\nu}v^{*\mu}\dot{v}^{*\nu}\,.$$ 
\end{abstract}
 {Keywords: \itshape Thomas precession, Wigner rotation, inverse problem of kinematics, proper acceleration, proper angular velocity, tetrad formalism.}

\begin{flushleft}
{\bf{Introduction}}
\end{flushleft}

In the previous paper \cite{1} the equations of the inverse problem of kinematics were derived. These equations follow from the dependence of the characteristics of the rigid reference frame s on the motion parameters, which in turn are largely based on the well-known transformation into a rigid non-inertial reference frame proposed by Nelson \cite{2}, \cite{3} (see also \cite{14}). This transformation is undoubtedly correct for the following reasons. First, the metric following from this transformation satisfies the principle of general form invariance \cite{4}, i.e. is precisely the metric of a rigid accelerated and rotating reference frame (see \cite{2}, \cite[p. 331, formula (13.71) corrected for the absence of space-time curvature in SRT]{5}. Secondly, the proper the acceleration and angular velocity of the non-inertial frame $s$ in this metric coincide with the independently computed kernel of the infinitesimal Lorentz transformation generator that relates two instantaneously comoving $s$ inertial reference frames at times $t$ and $t+\Delta t$ respectively, as specified in \cite[p. 138, problem 10.25]{6}.

However, although there is no reason to doubt the validity of this transformation, but since this transformation is still being studied, it may seem that the equations of the inverse problem of kinematics are not yet sufficiently substantiated. These equations themselves are necessary to determine the law of motion of a non-inertial frame of reference having given characteristics. Therefore, for more reliable confirmation, it is important to derive them in a different way. Such a method will be shown in section \ref{s}, but preliminary, in section \ref{f} of this paper, some useful formulas are given regarding the kinematics of rotations and, in particular, Wigner's proper rotation \cite{7}, \cite{8}, \cite[formula (20)]{9}, which arises in a moving reference frame when boosting from one laboratory frame to another.

The most general possible motion of a rigid non-inertial reference frame must be determined invariantly, in a 4-dimensional form. For example, for rectilinear motion of a uniformly accelerated reference frame, the condition of its uniform acceleration can be formulated as one equality - the constancy of the square of the 4-vector of acceleration $d\Lambda^{0i}/dt\cdot d\Lambda^{0}_{\,\,\ ,\,i}/dt$ , where $t$ is the proper time of origin, and $\Lambda^{0i}$ is the 4-velocity vector \cite[p. 22, problem to paragraph 7]{10}. For the most general curvilinear motion of an arbitrary rigid frame of reference, one such requirement is clearly not enough. These conditions must relate the components of 4-vectors, which the moving frame possesses, with its proper invariants of the given reference frame. Therefore, the equations for the 4-acceleration $d\Lambda^{0i}/dt=a^{i}$ and the 4-rotation tensor \cite[p. 174, formula (6.20)]{5}, which at first glance could be would use, in this case are useless. One of the required conditions (for a particular case of a reference frame rotating with the frequency of Thomas proper precession) was given in \cite[formula (5.29)]{11}. Completely invariant conditions of motion will be given in section \ref{s}. Looking ahead, we point out that these conditions are actually a requirement for the 4-scalar components of the characteristics of a given frame of reference. If these conditions are valid, then the form invariance of these equations under boost can be verified. The corresponding proof is given in section \ref{t}. The section \ref{four} discusses the difference between Thomas precession and Wigner rotation, as these concepts are sometimes confused with each other. In addition, in section \ref{five} we will consider the law of rotation of the reference frame's own tetrad, which was discussed in \cite[p. 171, formula (6.11)]{5}, \cite[formulas (4), (6)]{12}.

\subsubsection{Turn kinematics and Wigner rotation}\label{f}

Let us first recall some information from the kinematics of turns. When rotating the coordinate system around the unit vector $\mathbf{n}$, the components of the vector $\mathbf{r}(s)$ in the initial frame $s$ and the components of the same vector $\mathbf{r}(s')$ in the final frame system $s'$ are related by equality
\begin{equation} \label{1}
r^{\alpha }(s) =a^{\beta \alpha} r'^{\beta }(s')\,,\,\,\,\,\,\alpha,\beta=1,2,3,
\end{equation}
where matrix $a^{\beta\alpha}$ in coordinates: axis of rotation $\mathbf{n}$, angle of rotation $\phi$ has the form
\begin{equation} \label{2}
a^{\beta\alpha}=\delta^{\alpha\beta}\cos\phi+n^{\beta}n^{\alpha}(1-\cos\phi)-e^{\alpha\beta\gamma}n^{\gamma}\sin\phi\,.
   \end{equation}
 To avoid misunderstandings, we emphasize that hereinafter, a passive point of view on rotations is adopted, according to which the coordinate system rotates, while the vector itself remains unchanged. In addition, for any rotation matrix $a^{\beta\alpha}$, the following orthogonality relations hold:
 \begin{equation} \label{3}
a^{\beta \alpha} a^{\gamma \alpha} =a^{\alpha \beta} a^{\alpha\gamma} =\delta ^{\beta\gamma}
\end{equation}
 and the equalitys of "annihilation"
\begin{equation} \label{4}
 e^{\alpha \mu \nu } a^{\mu \beta } a^{\nu \gamma } =e^{\mu \beta \gamma } a^{\alpha \mu } ,\,\,\,\,
 e^{\alpha \mu \nu } a^{\beta \mu } a^{\gamma \nu } =e^{\mu \beta \gamma } a^{\mu \alpha }.
 \end{equation}
The rotation matrix $a^{\beta\alpha}(t)$ satisfies the equation
 \begin{equation} \label{5}
 \frac {da^{\alpha \beta}}{dt} =-e^{\alpha\mu\nu} \omega'^{\mu} a^{\nu \beta}\,,
 \end{equation}
 where $\boldsymbol{\omega'}$ is the angular velocity in the frame $s'$
 \begin{equation} \label{6}
 \omega'^{\nu }=\frac {1}{2}\,e^{\alpha\lambda\nu }a^{ \lambda\beta } \frac{da^{\alpha \beta } }{dt}\,\,.
\end{equation}

Let us now consider the motion of the inertial frame $s'$ with the speed $\mathbf{v}(t)$ relative to the laboratory frame $S$. If the inertial frame $s'$ is oriented in such a way that the transformation of the space-time coordinates from $S$ to $s'$ is a pure boost, then it is conventionally said that the frame $s'$ is oriented "without turn"\, relative to $S $. Now let's move on to the new laboratory reference frame $S^*$, which moves with the speed $\mathbf{u}$ "without turn" relative to $S$. It is well known that in the new laboratory reference frame origin  of the frame $s'$ moves with velocity (hereinafter, the system of units is used, in which $c=1$ )
\begin{equation} \label{7}
\mathbf{v^*}=\frac{\sqrt{1-u^{2} }\,\,\mathbf{v-u}}{1-\mathbf{uv}} +\frac{(1-\sqrt{1-u^{2} } )\mathbf{(uv)u}}{u^{2} (1-\mathbf{uv})}\,\,.
\end{equation}
However, it turns out that the transformation from $S^*$ to $s'$ is not a pure boost. In other words, this means that, in order to obtain the $s'$ coordinate system, the axes of the new reference frame $s$, which are oriented "without turn"\, relative to $S^*$ (which moves with the velocity  $\mathbf{v}(t)$ from \eqref{7}), it is  required to additionally rotate by some angle $\phi_W$ around some unit vector $\mathbf{n}$. A rather lengthy calculation, which we will skip here, states that this proper turn is (see \cite[formula (20)]{9}, \cite[formula (42)]{13})
\begin{equation} \label{8}
\mathbf{n}\,\tg \frac{\phi_W }{2}=\mathbf{n}\,\,\frac{\sin \phi_W }{1+\cos \phi_W }=\frac{\mathbf{u}\times \mathbf{v}}{\left( 1+\sqrt{1-{{u}^{2}}} \right)\left( 1+\sqrt{1-{{v}^{2}}} \right)-\mathbf{uv}}\,\,.
\end{equation}
This rotation should be called \textit{the Wigner rotation}. If $\mathbf{u}$ is small, then
\[\mathbf{n}\,\,\phi_W=\frac{1-\sqrt{1-v^2}}{v^2}\,\,\mathbf{u}\times \mathbf{v}.\]
Denote the Wigner rotation matrix by $b^{\beta\alpha}_W$ , so that \eqref{1} looks like
\begin{equation} \label{10}
r^{\alpha }(s) =b^{\beta \alpha}_{W} \,\,r'^{\beta }(s') ,
\end{equation}
For a small angle $\phi_W$ the rotation matrix \eqref{2} is
\begin{equation}\label{11}
  b^{\beta\alpha}_W=\delta^{\beta\alpha}-e^{\alpha\beta\gamma}n^{\gamma}\phi_{W}=\delta^{\beta\alpha}+\frac{1-\sqrt{1-v^2}}{v^2}(v^\alpha u^{\beta}-v^{\beta}u^{\alpha}).
\end{equation}
By definition \eqref{6}, the angular velocity of Wigner rotation is
\begin{equation} \label{12}
 \omega'^{\nu }_W=\frac {1}{2}\,\,e^{\alpha\mu\nu }\,b^{ \mu\beta }_W \,\frac{d\,b^{\alpha \beta }_W}{dt}\,\,.
\end{equation}
Substituting \eqref{11} here we get that it is equal to
\begin{equation}\label{13}
  \boldsymbol{\omega}'_W=\frac{d\,(\mathbf{n}\,\,\phi_W)}{dt}=\frac{(1-\sqrt{1-v^2})^2}{v^4\sqrt{1-v^2}}\,\mathbf{(v\dot{v})}\,\mathbf{u\times v}+\frac{1-\sqrt{1-v^2}}{v^2}\,\mathbf{u\times\dot{v}}.
\end{equation}

Let us now consider the general case of motion of an inertial frame, when it (we denote it by $k$) had a different orientation relative to the frame $S$ than the "without turn" \,frame $s'$. In other words, let the components of the vector $\mathbf{r}(s')$ in the initial frame $s'$ and the components of the same vector $\mathbf{r}(k)$ in the frame $k$ be related by the equality
\begin{equation}\label{14}
   r^\beta(s')=a^{\gamma\beta}r^{\gamma}(k).
\end{equation}
Substituting \eqref{14} into \eqref{10} we get that the resulting rotation matrix between the initial system $s$ and the final system $k$ is
\begin{equation}\label{15}
   a^{*\gamma\alpha}=a^{\gamma\beta}b^{\beta\alpha}_W \,.
\end{equation}

Thus, the parameters of the reference frame $k$, which moves with the speed $\mathbf{v}$ and has the orientation $a^{\gamma\alpha}$ with respect to $S$, with a pure boost with the velocity $\mathbf{u} $ are transformated according to the laws \eqref{7}, \eqref{15}.

\subsubsection{Invariant conditions for the motion of a non-inertial reference frame.}\label{s}

Consider now the following 4 values consisting of 4 components, where the first index of each character is responsible for its number, and the second - for its component:
\begin{equation}\label{16}
  \Lambda^{0i}=(\Lambda^{00},\Lambda^{0\alpha})=\left(\frac{1}{\sqrt{1-v^2}}\,,\frac{v^{\alpha}}{\sqrt{1-v^2}}\right),
\end{equation}

\begin{equation}\label{17}
  \Lambda^{\alpha i}=(\Lambda^{\alpha 0},\Lambda^{\alpha\beta})=\left(\frac{v^{\gamma}a^{\alpha\gamma}}{\sqrt{1-v^2}}\,,a^{\alpha\beta}+\frac{1-\sqrt{1-v^2}}{v^2\sqrt{1-v^2}}v^{\beta}v^{\mu}a^{\alpha\mu}\right),
\end{equation}
where $v^{\alpha}$ is the 3-vector of the  velocity of the non-inertial reference frame origin , $a^{\alpha\beta}$ is the rotation matrix, and the Greek indices, as usual, run through the values 1,2,3; Latin indices is 0,1,2,3. It is easy to check that these quantities satisfy the orthonormality relations
\begin{equation} \label{18}
	{{\Lambda }^{0i}}\Lambda ^{0}_{\ \ i}=1,\,\,\,{{\Lambda }^{0i}}\Lambda ^{\alpha }_{\ \ i}=0,\,\,\,{{\Lambda }^{\alpha i}}\Lambda ^{\beta }_{\ \ i}={-\,{\delta }^{\alpha \beta }}.
\end{equation}
 From a mathematical point of view, $\Lambda^{0i}$ , $\Lambda^{\alpha i}$ are coefficients in the general Lorentz transformation from some laboratory frame reference $S$: $(T,\mathbf{R})$ to inertial frame $s':(t,\mathbf{r'})$
 \[T=\Lambda^{\alpha 0}r'^{\alpha}+\Lambda^{00}t,\,\,\,\,\,\,R^{\beta}=\Lambda^{\alpha\beta}r'^{\alpha}+\Lambda^{0\beta}t.\]
If we consider $r'^{\alpha}$ and $t$ as parameters, then from the definition of $\Lambda^{\alpha i}=dX^{i}/dr'^{\alpha}$ and $\Lambda ^{0i}=dX^{i}/dt$ we see that they are 4-vectors. In addition, under any continuous transformation from the general Lorentz group, the quantities $\Lambda^{\alpha i}$ , $\Lambda^{0i}$ $(i=0,1,2,3)$ are transformed into quantities of the same mathematical forms \eqref{16}, \eqref{17}. As is well known, the zero vector $\Lambda^{0i}$ is the timelike 4-velocity vector of the beginning of $s'$. Its meaning lies in the fact that it is unit and tangent to the world line of the origin of the non-inertial frame $s'$. The 4-vector $\Lambda^{\alpha i}$ is a spacelike unit vector of the $\alpha$-th axis of the $s'$ coordinate system. Let's compose now products
\begin{equation} \label{19}
    {{{W}'}^{\gamma }}=\Lambda ^{0}_{\;\;i}\frac{d\Lambda ^{\gamma i}}{d{t}}\,\,,
\end{equation}
\begin{equation} \label{20}
	{{{\Omega }'}^{\gamma }}=-\frac{1}{2}\,{{e}^{\alpha \mu \gamma }}\Lambda _{\;\;i}^{\mu }\frac{d{{\Lambda }^{\alpha i}}}{d{t}}\,\,.
\end{equation}
In an instantaneously comoving inertial frame, these products are equal to $W'^{\gamma}=a^{\gamma\beta}\dot{v}^{\beta}$ and ${({\Omega }'}^{\gamma)}={{e}^{\,\alpha \mu \gamma }}a^{\mu\beta}\dot{a}^{\alpha\beta}/2$ , i.e. are their proper acceleration and their proper angular velocity of the frame $s'$, respectively. Due to the 4-vector nature of the quantities $\Lambda^{\alpha i}$ , $\Lambda^{0i}$, these equations will be valid in any reference frame. This means that the equations \eqref{19}, \eqref{20} are the desired kinematic, relativistically invariant conditions for the motion of a given non-inertial frame with known characteristics. Differentiate \eqref{16}, \eqref{17} and substitute these values into \eqref{20}. After expanding the brackets, renaming the indices, casting like terms, and using the equalities \eqref{4} and \eqref{6}, we finally get
\begin{equation}\label{21}
\Omega'^{\gamma}=a^{\gamma\alpha}\,\frac{1-\sqrt{1-v^{2}}}{v^{2}\sqrt{1-v^{2}}}\,\,e^{\alpha\mu\nu}v^{\mu}\dot{v}^{\nu}+\omega'^{\gamma}.
\end{equation}
Proceeding similarly, it is easy to get from \eqref{19}
\begin{equation}\label{22}
  W'^{\gamma}=a^{\gamma \alpha} \left[ \frac{\dot{v}^{\alpha}}{\sqrt{1-v^2}} +
 \frac{1-\sqrt{1-v^2}}{v^2(1-v^2)} (\mathbf{v\dot{v}})v^{\alpha}  \right ].
\end{equation}
The equations \eqref{21} and \eqref{22}, after replacing $v^{\alpha}=a^{\beta\alpha}v'^{\beta}$, reduce to the equations of the inverse problem of kinematics, as shown in \cite {1}.

\subsubsection{Checking the form invariance of the reference  frame  characteristics}\label{t}

For full confidence in the formula of proper acceleration and proper angular velocity \eqref{19}, \eqref{20} must be checked. By their meaning, these formulas are actually a requirement for the invariance of the characteristics of the reference frame $s'$ from boosts. Therefore, it can be written that in the new laboratory reference frame $S^*$ and in the old reference frame $S$ the proper acceleration and the angular velocity of the $s'$ frame must be equal when substituting \eqref{7}, \eqref{15}. Unfortunately, such a direct substitution into the equalities \eqref{21}, \eqref{22} (where all values are considered with asterisks) of the formulas \eqref{7}, \eqref{15} leads to cumbersome calculations. Meanwhile, we note that an arbitrary boost can be considered as a composition of a set of arbitrary infinitesimal boosts. Therefore, in order to show the validity of the equalities \eqref{21}, \eqref{22} in the general case, it suffices to prove their validity for the infinitesimal Lorentz transformation. For such a transformation, \eqref{7} shows that
\begin{equation}\label{25}
  \mathbf{v}^*=\mathbf{v}(1+\mathbf{uv})-\mathbf{u},\,\,\,\,\,\mathbf{\dot{v}}^*=\mathbf{\dot{v}}(1+\mathbf{uv})+\mathbf{v(u\dot{v})}.
\end{equation}
When substituting into an expression
\[W'^{\gamma}=a^{*\gamma\alpha}\left[\frac{\dot{v}^{*\alpha}}{\sqrt{1-v^{*2}} }+\frac{1-\sqrt{1-v^{*2}}}{v^{*2}(1-v^{*2})}(\mathbf{v^*\dot{v} ^*})v^{*\alpha} \right]\]
  \eqref{25} formulas, as well as \eqref{15} and \eqref{11}, after some algebraic transformations, we get that the mathematical form of $\mathbf{W}'$ has not changed
\[W'^{\gamma}=a^{\gamma \alpha} \left[ \frac{\dot{v}^{\alpha}}{\sqrt{1-v^2}}+
  \frac{1-\sqrt{1-v^2}}{v^2(1-v^2)} (\mathbf{v\dot{v}})v^{\alpha} \right ]. \]
 
 Similarly to the above, substituting \eqref{6} and \eqref{15} into \eqref{21}, differentiating and taking into account the orthogonality of the Wigner matrix, which is similar to the \eqref{3} equation, we get that
  \[\Omega'^{\gamma}=a^{*\gamma\alpha}\frac{1-\sqrt{1-v^{*2}}}{v^{*2}\sqrt{1-v^{*2}}}\,\,e^{\alpha\mu\nu}v^{*\mu}\dot{v}^{*\nu}+\omega'^{*\gamma}=\]
  \[=(a^{\gamma\beta}b^{\beta\alpha}_{W})\frac{1-\sqrt{1-v^{*2}}}{v^{*2}\sqrt{1-v^{*2}}}\,\,e^{\alpha\mu\nu}v^{*\mu}\dot{v}^{*\nu}+\frac{1}{2}\,\,e^{\alpha\mu\gamma}(a^{\mu\nu}b^{\nu\lambda}_{W})\frac{d}{dt}(a^{\alpha\beta}b^{\beta\lambda}_{W})=\]
 \[=(a^{\gamma\beta}b^{\beta\alpha}_{W})\frac{1-\sqrt{1-v^{*2}}}{v^{*2}\sqrt{1-v^{*2}}}\,\,e^{\alpha\mu\nu}v^{*\mu}\dot{v}^{*\nu}+\frac{1}{2}\,\,e^{\alpha\mu\gamma}\left[a^{\mu\nu}a^{\alpha\beta}b^{\nu\lambda}_{W}\dot{b}^{\beta\lambda}_{W}+a^{\mu\nu}\dot{a}^{\alpha\beta}(b^{\nu\lambda}_{W}b^{\beta\lambda}_{W})\right]=\]
 \[=(a^{\gamma\beta}b^{\beta\alpha}_{W})\frac{1-\sqrt{1-v^{*2}}}{v^{*2}\sqrt{1-v^{*2}}}\,\,e^{\alpha\mu\nu}v^{*\mu}\dot{v}^{*\nu}+\frac{1}{2}\,\,(e^{\alpha\mu\gamma}a^{\mu\nu}a^{\alpha\beta})b^{\nu\lambda}_{W}\dot{b}^{\beta\lambda}_{W}+\frac{1}{2}\,\,e^{\alpha\mu\gamma}a^{\mu\nu}\dot{a}^{\alpha\beta}\delta^{\nu\beta}.\]
Further, in the second term, we take into account the first of the "annihilation"\, equalities \eqref{4} and after that the equality \eqref{12}
\[\Omega'^{\gamma}=(a^{\gamma\beta}b^{\beta\alpha}_{W})\frac{1-\sqrt{1-v^{*2}}}{v^{*2}\sqrt{1-v^{*2}}}\,\,e^{\alpha\mu\nu}v^{*\mu}\dot{v}^{*\nu}+\frac{1}{2}\,\,e^{\tau\beta\nu}a^{\gamma\tau}b^{\nu\lambda}_{W}\dot{b}^{\beta\lambda}_{W}+\frac{1}{2}\,\,e^{\alpha\mu\gamma}a^{\mu\beta}\dot{a}^{\alpha\beta}=\]
 \[=(a^{\gamma\beta}b^{\beta\alpha}_{W})\frac{1-\sqrt{1-v^{*2}}}{v^{*2}\sqrt{1-v^{*2}}}\,\,e^{\alpha\mu\nu}v^{*\mu}\dot{v}^{*\nu}+\omega'^{\tau}_{W}a^{\gamma\tau}+\frac{1}{2}\,\,e^{\alpha\mu\gamma}a^{\mu\beta}\dot{a}^{\alpha\beta}=\]
\begin{equation}\label{26}
=a^{\gamma\beta}\left[b^{\beta\alpha}_{W}\frac{1-\sqrt{1-v^{*2}}}{v^{*2}\sqrt{1-v^{*2}}}\,\,e^{\alpha\mu\nu}v^{*\mu}\dot{v}^{*\nu}+\omega'^{\beta}_{W}\right]+\frac{1}{2}\,\,e^{\alpha\mu\gamma}a^{\mu\beta}\dot{a}^{\alpha\beta}.
\end{equation}
Consider separately the expression on the right side \eqref{26} in square brackets. Substituting \eqref{11}, \eqref{13}, \eqref{25} here we get after some calculations that
\[b^{\beta\alpha}_{W}\frac{1-\sqrt{1-v^{*2}}}{v^{*2}\sqrt{1-v^{*2}}}\,\,e^{\alpha\mu\nu}v^{*\mu}\dot{v}^{*\nu}+\omega'^{\beta}_{W}=\frac{1-\sqrt{1-v^{2}}}{v^{2}\sqrt{1-v^{2}}}\,\,e^{\beta\mu\nu}v^{\mu}\dot{v}^{\nu}+\frac{(1-\sqrt{1-v^{2}})^2}{v^{4}\sqrt{1-v^{2}}}\,\,k^{\beta},\]
where the vector $\mathbf{k}$ whose component is on the right side of this equality is equal to
\[\mathbf{k}=v^2\mathbf{\dot{v}\times u}+\mathbf{(uv)}\mathbf{v\times\dot{v}}-\mathbf{(v \dot{v})}\mathbf{v\times u}+\left[\mathbf{\dot{v}(v\times u)}\right]\mathbf{v}.\]
We notice, that
\[\mathbf{k\dot{v}=kv=ku}=0.\]
If the vectors $\mathbf{\dot{v}}$, $\mathbf{v}$ , $\mathbf{u}$ do not lie in the same plane, then this is possible only if the vector $\mathbf{ k}$ is identically equal to zero. If these vectors are coplanar, then $\mathbf{\dot{v}(v\times u)}=0$. In this case, all the remaining vector terms in $\mathbf{k}$ lie on the same axis, and the length of the vector $\mathbf{k}$ can be calculated by the definition of the vector and scalar product. Let the angle between $\mathbf{\dot{v}}$ and $\mathbf{u}$ be $\varphi_1$ , and between $\mathbf{v}$ and $\mathbf{u}$ be $\varphi_2$ . Then choosing positive direction of the vector $\mathbf{v\times\dot{v}}$ we have
\[|\mathbf{k}|=v^2\dot{v}u\left[-\sin\varphi_1+\cos\varphi_2\sin(\varphi_1+\varphi_2)-\sin\varphi_2\cos(\varphi_1+\varphi_2)\right]=0.\]
Therefore, in any case $\mathbf{k}=0$ and
\begin{equation}\label{27}
  b^{\beta\alpha}_{W}\frac{1-\sqrt{1-v^{*2}}}{v^{*2}\sqrt{1-v^{*2}}}\,\,e^{\alpha\mu\nu}v^{*\mu}\dot{v}^{*\nu}+\omega'^{\beta}_{W}=\frac{1-\sqrt{1-v^{2}}}{v^{2}\sqrt{1-v^{2}}}\,\,e^{\beta\mu\nu}v^{\mu}\dot{v}^{\nu}.
\end{equation}
Thus, it follows from here that the angular velocity \eqref{26} is form-invariant \[\Omega'^{\gamma}=a^{*\gamma\alpha}\frac{1-\sqrt{1-v^{*2}}}{v^{*2}\sqrt{1-v^{*2}}}\,\,e^{\alpha\mu\nu}v^{*\mu}\dot{v}^{*\nu}+\frac{1}{2}\,\,e^{\alpha\mu\gamma}a^{*\mu\lambda}\dot{a}^{*\alpha\lambda}=\]
\[=a^{\gamma\alpha}\frac{1-\sqrt{1-v^{2}}}{v^{2}\sqrt{1-v^{2}}}\,\,e^{\alpha\mu\nu}v^{\mu}\dot{v}^{\nu}+\frac{1}{2}\,\,e^{\alpha\mu\gamma}a^{\mu\lambda}\dot{a}^{\alpha\lambda}.\]

\subsubsection{On the relationship between Thomas precession and Wigner rotation}\label{four}

Let in some laboratory frame $S$ the origin of the rigid reference frame $s'$ move relativistically translational ($a^{*\beta\alpha}=\delta^{\beta\alpha}$ , i.e. "without turn" with respect $S$) with the speed $\mathbf{v}$ . This frame of reference has its proper angular velocity equal to the Thomas precession frequency
 \[\mathbf{\Omega}_{T}=\frac{1-\sqrt{1-v^{2} } }{v^{2} \sqrt{1-v^{2} } }\,\,\mathbf{v\times \dot{v}}.\]
Let's make successive (at close moments of time) boosts to instantaneously comoving $s'$ inertial reference frames. In this case, the new Thomas precession of the $s'$ frame (the first term on the left side of \eqref{27}) will be small (because of the low relative velocity) and the entire Thomas proper precession will be attributed to the Wigner rotation. This conclusion is not difficult to prove. Indeed, let at the moment $T+\Delta T$ the velocity of the frame $s'$ relative to $S$ becomes $\mathbf{v}+\Delta \mathbf{v}$. Let's move to a reference frame moving with the velocity $\mathbf{u=v}$. By making a substitution in \eqref{8} 
\[\mathbf{u}\rightarrow\mathbf{v},\,\,\,\,\,\mathbf{v}\rightarrow\mathbf{v}+\Delta\mathbf{v}\]
and taking into account that the Wigner rotation angle $\phi_{W}$ is small, as a result we get that it is equal to   \[\mathbf{n}\,\phi_{W} =\frac{1-\sqrt{1-v^{2}}}{v^{2}\sqrt{1-v^{2}}}\,\mathbf{v}\times \Delta \mathbf{v}.\]
 Obviously, this angle coincides with the natural frequency of the Thomas precession multiplied by the proper time interval. This circumstance is sometimes regarded as confirmation that Thomas precession and Wigner rotation are different names for the same physical phenomenon. This opinion is, of course, erroneous, since for an arbitrary (and even for a very small) boost, it is absolutely necessary to distinguish between Wigner rotation and Thomas precession.

This circumstance can also be explained using the equation \eqref{27} as an example, if you boost from $S$ not to an instantaneously comoving reference frame, but to an arbitrary one with a velocity of $\mathbf{u}$ . In the new laboratory reference frame $S^*$ the origin $s'$ already has the velocity $\mathbf{v^{*}}$ , which is related to $\mathbf{v}$ and $\mathbf{u}$ by the formula \eqref{7}, and the coordinate axes $s'$ experience an additional Wigner rotation \eqref{8}. Consequently, the angular velocity of proper rotation $s'$ in the new system is added according to the formula \eqref{27} from two components: the new Thomas proper precession (the first term on the left side) and the Wigner rotation (the second term on the left side).

Thus Wigner proper rotation should not be confused with Thomas proper precession; although they are closely related, they are not equivalent concepts. The essence of Thomas's discovery is that if a rigid reference frame $s'$ moves in such a way that at each moment of laboratory time its coordinate axes coincide with the axes of the inertial reference frame  oriented "without turn"\, with respect to the laboratory frame, then the reference frame $s'$ has its proper precession, which is the Thomas precession. In addition to the Thomas precession, the proper Wigner rotation, as can be seen from the \eqref{27} equation, is determined by the $b^{\alpha\beta}$ matrix and appears for any boost.

In the case when the reference frame $s'$ in the original laboratory frame moved rectilinearly and uniformly acceleratly without its proper rotation, then its proper Thomas precession is equal to zero. Then it follows from \eqref{27} that in the new laboratory frame, the Wigner rotation and the Thomas precession of the $s'$ frame must be opposite and compensate each other. This assertion will be independently verified in another paper.

\subsubsection{The motion law of proper tetrad}\label{five}

Let us now find the motion law of proper tetrad. If we differentiate the components of $\Lambda^{\alpha 0}$ from \eqref{17}, replace the derivatives of $\dot{a}^{\alpha\beta}$ in the resulting expression according to the \eqref{5} equation, and  take into account expression for $\omega'^{\alpha}$ from \eqref{21}, we get that
\[\frac{d\Lambda^{\alpha 0}}{dt}=\frac{a^{\alpha\nu}\dot{v}^{\nu}}{\sqrt{1-v^2}}-\frac{1-\sqrt{1-v^2}}{v^2(1-v^2)}(e^{\alpha\mu\nu}a^{\nu\gamma}a^{\mu\lambda})e^{\lambda\psi\varphi}v^{\gamma}v^{\psi}\dot{v}^{\varphi}+\]
\[+\frac{a^{\alpha\nu}v^{\nu}(\mathbf{v \dot{v}})}{\sqrt{1-v^2}^3}+e^{\alpha\beta\gamma}\Omega'^{\gamma}\frac{v^{\nu}a^{\beta\nu}}{\sqrt{1-v^2}}\,\,.\]
Further, everywhere we substitute instead of the derivatives $\dot{v}^{\nu}$ their values according to the equation
\[\dot{v}^{\varphi}=\sqrt{1-v^2}a^{\lambda\varphi}W'^{\lambda}-\frac{\sqrt{1-v^2}(1-\sqrt{1-v^2})}{v^2}(v^{\alpha}a^{\beta\alpha}W'^{\beta})v^{\varphi},\]
which follows from \eqref{22} and take into account the "annyhilation" equality \eqref{4}. After all the calculations, we get that
\begin{equation}\label{28}
  \frac{d\Lambda^{\alpha 0}}{dt}=W'^{\alpha}\frac{1}{\sqrt{1-v^2}}+e^{\alpha\beta\gamma}\Omega'^{\gamma}\frac{v^{\nu}a^{\beta\nu}}{\sqrt{1-v^2}}\,\,.
\end{equation}
Proceeding similarly, for the derivative component $\Lambda^{\alpha\mu}$ from \eqref{17} one can find its following value
\begin{equation}\label{29}
  \frac{d\Lambda^{\alpha\mu}}{dt}=W'^{\alpha}\frac{v^{\nu}a^{\alpha\mu}}{\sqrt{1-v^2}}+e^{\alpha\beta\gamma}\Omega'^{\gamma}a^{\beta\mu}+e^{\alpha\beta\gamma}\Omega'^{\gamma}\frac{1-\sqrt{1-v^2}}{v^2\sqrt{1-v^2}}v^{\mu}v^{\nu}a^{\beta\nu}.
\end{equation}
The equations \eqref{28} and \eqref{29} are written in a unified form as
\begin{equation}\label{30}
  \frac{d\Lambda^{\alpha i}}{dt}=W'^{\alpha}\Lambda^{0 i}+e^{\alpha\beta\gamma}\Omega'^{\gamma}\Lambda^{\beta i}.
\end{equation}
Similarly to the previous calculations, differentiating the components \eqref{16} we get that
\begin{equation}\label{31}
  \frac{d\Lambda^{0 i}}{dt}=W'^{\alpha}\Lambda^{\alpha i}.
\end{equation}
Multiplying \eqref{30}, \eqref{31} by $\Lambda^{\mu}_{i}$ and using the orthogonality property \eqref{18}, one can check the validity of the equalities \eqref{19}, \eqref{20}.

\subsubsection*{Conclusion}

The purpose of this paper was to show that the equations of the inverse problem of kinematics are valid in themselves, regardless of the truth of the Lorentz-M{\o}ller-Nelson transformation \cite{2}, \cite{3}. The way to achieve this purpose was the formulation (in sec. 2) of the invariant conditions of motion \eqref{19}, \eqref{20} for an arbitrary rigid non-inertial reference frame  with specified characteristics. It turned out that the equations of the inverse kinematics problem  are a direct consequence of these conditions. Checking \eqref{19}, \eqref{20} for small boosts (sec. 3) it was shown (sec. 4) that if in one laboratory frame the non-inertial rigid frame $s'$ had only its proper Thomas precession, then in the new a laboratory reference frame, this a frame $s'$ would have a combination \eqref{27} of two rotations: the new Thomas proper precession and the Wigner rotation, which together give the original Thomas precession frequency.

Also in the paper the motion law of proper tetrad \eqref{30}, \eqref{31} was obtained. These formulas are useful, for example, when discussing the motion of spin particles in a field from the point of view of the relativity theory.

\begin{flushleft}
{\bf{Gratitude}}
\end{flushleft}

The author is very grateful to Professor N. G. Migranov for fruitful discussions and support.

\textbf{After paper.} I add some links to paper. Also I want to illustrate section 3 with the following words, which may be better understood. In the general case, upon transition to any reference frame, we have \eqref{27}
$$ \frac{1-\sqrt{1-v^{2}}}{v^{2}\sqrt{1-v^{2}}}\,\,e^{\beta\mu\nu}v^{\mu}\dot{v}^{\nu}=\omega'^{\beta}_{W} + b^{\beta\alpha}_{W}\,\frac{1-\sqrt{1-v^{*2}}}{v^{*2}\sqrt{1-v^{*2}}}\,\,e^{\alpha\mu\nu}v^{*\mu}\dot{v}^{*\nu}\,.$$
or in other words 

\[\textbf{old Thomas's precession}\Longrightarrow \,\,\textbf{new Wigner's rotation}\, +  \]
\[+\,\textbf{new Thomas precession turned  by the Wigner angle}\,.\]

For the particular case of transition to an instantaneously comoving frame of reference we have
$$ \frac{1-\sqrt{1-v^{2}}}{v^{2}\sqrt{1-v^{2}}}\,\,e^{\beta\mu\nu}v^{\mu}\dot{v}^{\nu}=\omega'^{\beta}_{W} \,.$$
or in other words 

\[\textbf{old Thomas's precession}\Longrightarrow \,\,\textbf{new Wigner's rotation}\,.\]

Finally, it is also possible that the old precession of Thomas was equal to zero (when the reference frame moves in a straight line and accelerated). Then we have that
$$ \omega'^{\beta}_{W} + b^{\beta\alpha}_{W}\,\frac{1-\sqrt{1-v^{*2}}}{v^ {*2}\sqrt{1-v^{*2}}}\,\,e^{\alpha\mu\nu}v^{*\mu}\dot{v}^{*\nu}= 0\,$$
or in other words, Wigner rotation and new Thomas precession cancel each other out
\[\textbf{new Wigner's rotation}\, +  \]
\[+\,\textbf{new Thomas precession turned  by the Wigner angle}=0\,.\]
We will prove this compensation in another paper.

\end {document}